\documentclass[aps,twocolumn,amsmath,amssymb,floatfix,superscriptaddress]{revtex4-2}

\usepackage{bm,graphicx,url,epsf,color}
\usepackage{wasysym}
\usepackage{MnSymbol}
\usepackage[normalem]{ulem}
\usepackage{natbib}
\usepackage{gensymb}
\usepackage{placeins}
\usepackage{float}
\usepackage{hyperref}
\usepackage{cleveref}
\usepackage{natbib}
\usepackage{amsmath}

\begin{document}
\title{\color{blue}\color{black}Strain-induced disorder from Schottky gates on high mobility 2DEGs
\color{black}}

\author{Sambit Mohapatra}
\affiliation{Universite Paris Saclay, CNRS, Centre de Nanosciences et de Nanotechnologies, Palaiseau, France}
\author{Lukas Berg}
\affiliation{Universite Paris Saclay, CNRS, Centre de Nanosciences et de Nanotechnologies, Palaiseau, France}
\author{Jean-Baptiste Cuvellier}
\affiliation{Universite Paris Saclay, CNRS, Centre de Nanosciences et de Nanotechnologies, Palaiseau, France}
\author{Yan Sun}
\affiliation{Universite Paris Saclay, CNRS, Laboratoire de physique des Solides, Orsay, France}
\author{Abdelhanin Aassime}
\affiliation{Universite Paris Saclay, CNRS, Centre de Nanosciences et de Nanotechnologies, Palaiseau, France}
\author{Alan Durnez}
\affiliation{Universite Paris Saclay, CNRS, Centre de Nanosciences et de Nanotechnologies, Palaiseau, France}
\author{Antonella Cavanna}
\author{Dominique Mailly}
\author{Ulf Gennser}
    \email[Correspondence email address: ]{ulf.gennser@cnrs.fr}
\affiliation{Universite Paris Saclay, CNRS, Centre de Nanosciences et de Nanotechnologies, Palaiseau, France}

\date{\today} 

\begin{abstract}
\color{blue}\color{black}Electrostatic gates on two-dimensional electron gases (2DEGs) in GaAs/AlGaAs heterostructures are essential for numerous studies of correlation driven phenomena, many performed within the fractional quantum Hall effect. However, the most commonly used type of gates, the metallic Schottky gate, has the propensity to adversely impact the electron mobility and suppress the fractional quantum Hall states. Here we demonstrate through commensurability oscillations under specially designed  nanostructured superlattice gates, that strain effects are responsible for the mobility reduction. The experimental results agree well with COMSOL  Multiphysics simulations. In a regular, unpatterned gate the disordered, granular nature of the metal film fills the same role as the superlattice in creating a strain field penetrating down to the 2DEG, as we show with a simplified model. These results will help in understanding, improving and finding novel methods to reduce strain-induced disorder in gated high-mobility 2DEGs. \color{black}
\end{abstract}


\maketitle

\section {Introduction} \label{sec:outline}
The fractional quantum Hall (FQH) regime, where energy gaps emerge at partly filled Landau levels due to electron-electron interactions, is a unique playground for correlation driven phenomena in condensed matter.  The recent demonstrations of anyonic statistics \cite{nakamura2020direct, bartolomei2020collider} has given further impetus to the investigations of the numerous FQH states \cite{glidic2023cross, glidic2024signature, veillon2024observation, ruelle2023comparing, lee2023partitioning, kundu2023anyonic, dutta2022distinguishing}.  The most common platform for the FQH effect are high-purity GaAs/AlGaAs-based structures containing two-dimensional electron gases (2DEGs), because of the possibility of achieving ultra-high electron mobilities and realizing exotic fractional states. So, for example, it has been possible to observe Jain states up to $\nu = 16/31$ \cite{chung2021ultra}, 6-flux states \cite{huang2024evidence}, and a number of even-denominator, possibly non-abelian states \cite{willett1987observation, eisenstein1992new, suen1992observation, luhman2008observation,wang2022even, wang2023next}. \par
Mesoscopic investigations have been very successful using the manipulation of local electric potentials and Coulomb interactions in devices based on 2DEGs in the integer quantum Hall regime. It has been possible to develop an 'electron quantum optics', which like its photonic counterpart relies on the manipulation of single particle states and the measurements of current-current correlations \cite{bocquillon2014electron}. Clearly, it is of large interest to implement this concept also in the FQH regime, in order to investigate and exploit the different facets of anyonic statistics.  However, it has been found to be much more difficult to realize the different parts such as single anyon emitters \cite{ruelle2024time-resolved} and different anyon interferometers \cite{nakamura2020direct, kundu2023anyonic}. \par
A prevalent issue is the notable suppression of FQH states in GaAs/AlGaAs-based 2DEGs in the presence of metallic Schottky gates used to regulate carrier density.\color{black} (An example of the effect of the gates on FQH states can be seen in supplementary \cref{fig:fig s1}.)
Understanding the underlying mechanisms responsible for this suppression and developing mitigation strategies are essential to fabricate electrostatic gates that do not appreciably  weaken FQH states. 
It has been suggested that the primary cause is the presence of interfacial strain at \color{black} the metal/semiconductor interface arising from different thermal expansion coefficients $\alpha$ of the materials\color{blue} 
\color{black} \cite{veillon2024observation, davies1994theory}. 
In fact, $\alpha$ of the typical gate metals are several times larger than that of GaAs. (For the temperature dependence of $\alpha$ for Al, Au and GaAs, see e.g. \cite{schmiedeshoff2006versatile, pamato2018thermal, loureno2004thermal}.) 
Due to the piezoelectric nature of GaAs this strain may create a potential modulation at the depth of the 2DEG.\par\color{cyan}\color{black}
There are essentially three mitigation strategies to limit gate-induced mobility reduction: reducing the strain by increasing the distance to the 2DEG or reducing the thickness of the metal; making specially designed gates; or by introducing an intermediate oxide layer. For example, it has been shown in several studies that this problem can be circumvented by burying the 2DEG deep inside the AlGaAs heterostructure \cite{singh2024topological} - though this limits the possibilities of realising nanostructures through quantum point contacts - or by introducing an intermediate layer of HfO\textsubscript{2}  \cite{dutta2022distinguishing, dutta2022isolated,veillon2024observation}.  However, a quantitative study of the strain-induced disorder in metallic Schottky gated FQH devices is missing, impeding the optimal design of gates in FQH studies. \par

By experimentally investigating gated Hall bar devices with varying thicknesses of metallic gates on Si-doped 2DEG systems, it is possible to probe how disorder is gradually introduced into the 2DEG with increasing thickness of the metal. 
Furthermore, by using periodic \color{black} nanostructured gates that create a superlattice strain potential in the heterostructure with superlattice periods of the same order of magnitude as typical grain structures in a deposited Al layer, we not only establish that at low temperature strain is introduced by the Schottky gates, but this also provides a way to quantify the strain potential. The modelling of the strain potentials give good agreement with the experimental results, and allows us to \color{blue}\color{black} propose specially designed nanostructured superlattice gates to minimise strain induced disorder.\color{black}

 \color{black}
 
\section {Samples} \label{sec:samples}
The devices used to study the effect of the global metallic gates are based on a two-dimensional electron gas at the interface of a GaAs/AlGaAs heterojunction at a distance of $\approx 100\,\mathrm{nm}$, from the surface. 
In the case of the samples with Schottky gates of uniform thickness (see \cref{fig:fig 2}a) \color{black} 
the 2DEG has a charge density $n_{s} = (8.5 \pm 0.2) \times 10^{14}\,\mathrm{m^{-2}} $ and a mobility $\mu = (1.25 \pm 0.2) \times 10^{2}\,\mathrm{m^{2} V^{-1} s^{-1}} $ measured at $6\,\mathrm{K}$. A uniform layer is used as the metallic gate.  
Simple Hall bars of typical width of $200\,\mathrm{\mu m}$
were fabricated by e-beam lithography, and the gate metals were deposited on the substrate at room temperature using electron beam evaporation deposition method \color{green}\color{black}at pressures around $1 \times 10^{-8}\,\mathrm{mbar}$.\color{black} 

In the case of nanopatterned Schottky gates (see below and \cref{fig:fig 2}b) \color{black} a higher-density 2DEG structure was used ($n_{s} = (2.8 \pm 0.1) \times 10^{15}\,\mathrm{m^{-2}}$, $\mu = (1.4 \pm 0.4) \times 10^{2}\,\mathrm{m^{2} V^{-1} s^{-1}} $). When Ti/Au was used as the patterned gate metals, a thin uniform metal layer was first deposited on the whole area, and then in a second fabrication step, using 
electron beam lithography \color{black} with standard positive e-beam resist and the lift-off method, the metallic array was realized.  In the case of Al-based gates, this sequence was not possible to use because of the oxidation of the Al surface in between depositions, and the order was inversed, so that the all-covering thin metal layer was deposited on top of the nanopatterned metallic array. 
We note that we did not observe any difference between the results for the Al and for the Ti/Au gates.\color{black}

Finally, we note that all gates were fabricated using e-beam evaporation directly onto the semiconductor surface kept at room temperature. 

\section{Results} \label{sec:Results}
\begin{figure}[t]
    \includegraphics[width=0.5\textwidth]{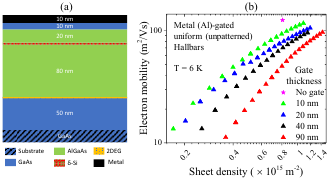}
\caption{\textbf{}(a) Schematic cross-sectional view of a 2DEG device with metallic gate. (b) Mobility vs. density characteristics measured at $6\,\mathrm{K}$ of devices with \color{blue}\color{black} a uniform (unpatterned) Al gate \color{black} for thicknesses between $10\,\mathrm{nm}$ and $90\,\mathrm{nm}$.  The magenta star represents the measurement on the ungated reference Hallbar from the same wafer.}
\label{fig:fig 1}

\end{figure}

We first illustrate how the thickness of a metal gate affects the classical low-temperature electronic transport properties.
In \cref{fig:fig 1}, density-mobility relationships at $6\,\mathrm{K}$ of a 2DEG in a gated Hallbar device are depicted for four gate thicknesses.  
Due to carrier screening of background Coulombic disorder, the mobility of 2DEGs is generally highly dependent on the carrier density. In order to identify how the gate thickness affects the mobility, it is therefore necessary to compare different samples for the same density, as regulated by the gate voltage. \color{black}
We observe that already a thin gate of $10\,\mathrm{nm}$ thickness significantly reduces the mobility compared to the nominal mobility measured on an ungated section of the same device (indicated by a star in the figure). Increasing gate thickness further lowers the density-mobility curves, indicating increased disorder with increased metal thickness. 


\begin{figure}[t]
    \includegraphics[width=0.5\textwidth]{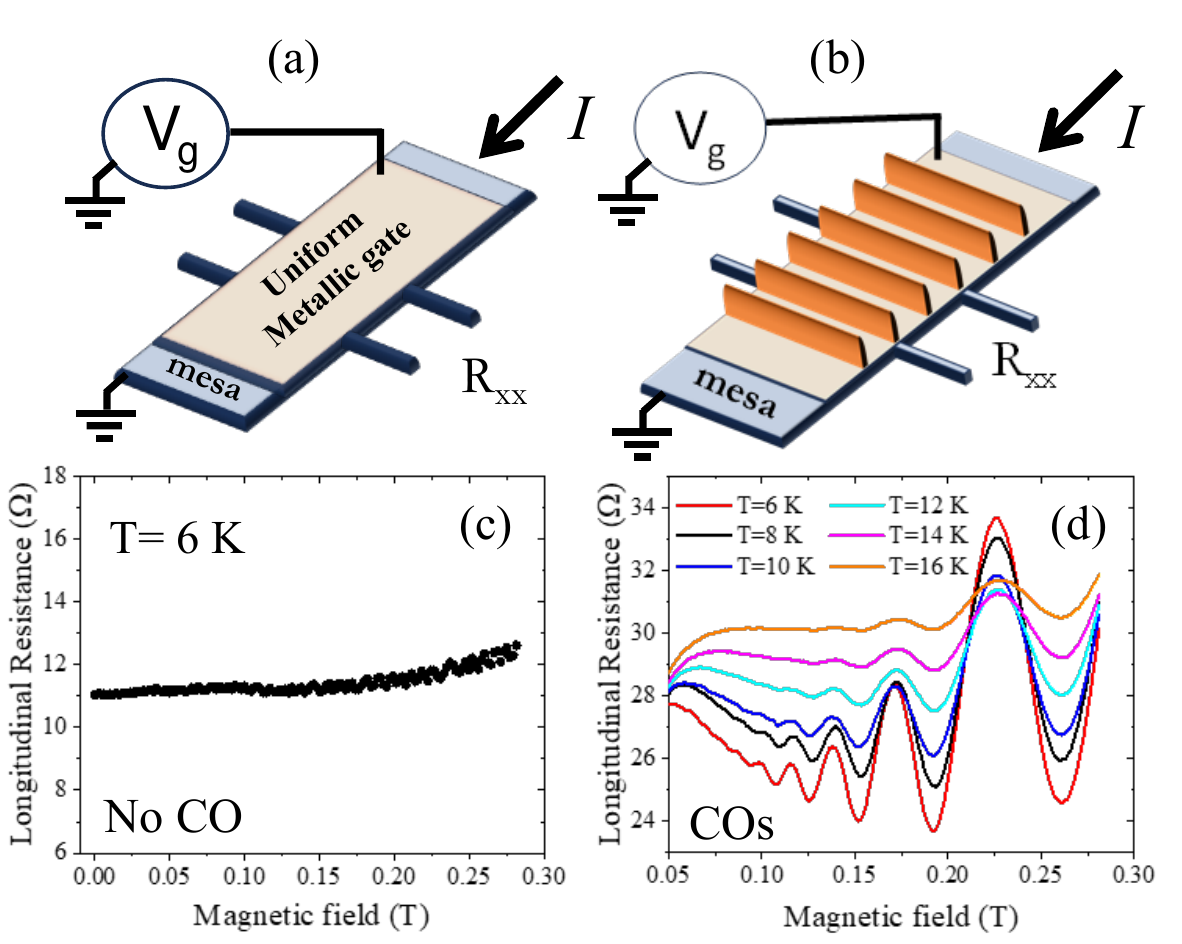}
\caption{Comparison of the longitudinal resistance between the uniform (unpatterned) and periodically patterned metallic gated Hallbar devices. (a) and (b) show the device schematics for the uniform metallic (Al) gate and the patterned gate (Al), respectively. $V_g$ represents the gate voltage applied to the uniform layer. (c) and (d) show the corresponding longitudinal resistance measurements for the two devices shown above them. The measurement shown in (c) is taken at a fixed sheet density $V_g=0\,\mathrm{V}$ at $T=6\,\mathrm{K}$ and the measurements in (d) are taken at a temperature of $T=6\,\mathrm{K}$ and at a fixed 2DEG density ($n_{s} = 2.81 \times 10^{15}\,\mathrm{m^{-2}}$) but for a different cooldown. The periodically patterned gate has a period of $241\,\mathrm{nm}$ for the array and a width of $90\,\mathrm{nm}$ for each bar of the array. The uniform layer (for both (c) and (d)) and the bar height of the periodic array are $20\,\mathrm{nm}$ and $130\,\mathrm{nm}$ of Al, respectively. Note that due to larger average metal height for the periodically patterned gate, the mobility is reduced with respect to the thin, unpatterned gate. The background subtracted normalized magnetoresistance data are shown in supplementary \cref{fig:fig s5}.}
\label{fig:fig 2}
\end{figure}

The mobility dependence on the gate thickness is an indication that the increased disorder is not related to defects on the metal-semiconductor interface (which would not increase with the metal thickness), but is rather due to strain effects. To explore the latter \color{black} in more detail, strain-induced lateral superlattices (LSL) can be a useful tool. 
In these, because of the piezoelectric nature of GaAs, the spatially varying strain produces a modulation in the electronic potential which can be experienced by the electrons in the 2DEG. This periodic potential \color{black} can lead to Weiss or commensurability oscillations (COs) in the magnetoresistance $\rho_{xx}(B)$, with an amplitude that depends on the strain.\cite{weiss1989magnetoresistance, davies1994theory, skuras1997anisotropic} 
However, for our purposes, it is necessary to ensure that the electrical potential remains constant over the whole gate area.  This is achieved by a 
special \color{black} metallic gate design, comprising a periodically patterned thick metallic array (under which the strain will be larger) atop a thinner uniform metallic layer (see \cref{fig:fig 2}b).
Low-temperature magnetoresistance observations on such devices reveal resistivity oscillations at low magnetic fields $B$ (\cref{fig:fig 2}d) for a fixed gate voltage, absent in the case of a regular unpatterned gate (\cref{fig:fig 2}c).  These are periodic in $1/B$, but differs from Shubnikov-de Haas (SdH) oscillations both from its smaller frequency and its lower magnetic field onset.
(For a comparison between these oscillations and the SdH oscillations, \color{green}\color{black} see supplementary \cref{fig:fig s3}a).\color{black}  The minima positions of these oscillations at a fixed density fits very well \color{green}\color{black}(see  supplementary \cref{fig:fig s3}b, c) \color{black} with the condition of commensurability between the period of the patterned array and the cyclotron orbit of the electrons, 
\begin{equation}
\frac{2R_{c}}{P} = i-\frac{1}{4},
\end{equation} 
where $R_{c}=\frac{{\hbar}k_F}{eB}$ is the cyclotron radius, $i$ is a positive integer, and $k_F=(2\pi n)^{1/2}$ with $n$ the electron density. $P$ is the period of the lateral superlattice.  The experimentally determined value from eq. 1 ($P=241 \pm 0.1\,\mathrm{nm}$, \color{blue}\color{black} supplementary \cref{fig:fig s3}c \color{black})  agrees very well with the lithographical value determined from scanning electron microscopy image ($P=242.8\,\mathrm{nm}$, \color{blue}\color{black} supplementary \cref{fig:fig s3}d), \color{black} confirming that these oscillations are COs \cite{weiss1989magnetoresistance, winkler1989landau, beenakker1989guiding}. The uniform metallic layer with uniform electric field, devoid of COs, (\cref{fig:fig 2}c) indicates the presence of the periodic strain potential originating from the additional periodic thicker metallic array.\par
Furthermore, the amplitude of the oscillations gives us a measure of the strain imposed on the 2DEG. 
As shown in \color{blue}\color{black} supplementary \cref{fig:fig s4} \color{black}, the oscillatory component of the COs fits very well with the asymptotic expression \cite{beton1990magnetoresistance,peeters1992electrical,mirlin1998weiss}
\begin{equation}
\frac{\rho^{osc}}{\rho_{0}} = A(\frac{\pi}{\omega_{c} \tau_{CO}})A(\frac{T}{T_{b}})\frac{(V_0/E_F)^2}{2 P\color{black}}\mu BL\sin(2\pi\frac{2R_{c}}{P\color{black}}), 
\end{equation} where $L=\frac{{\hbar}\mu k_F}{e}$
is the mean free path, $\mu$ is the mobility, and the 
first and second \color{black} functions $A(x)=\frac{x}{\sinh x}$ expresses the damping due to the CO lifetime and temperature damping, respectively. In the first, $\omega_c$ is the cyclotron frequency and $\tau_{CO}$ is the CO lifetime. In the second, $T_b = \frac{\omega_c{\hbar}P\color{black} k_F}{4\pi^2k_B}$ is the characteristic temperature \cite{endo2000envelope}
, and $k_B$ is the Boltzmann constant\color{black}. The relative 
amplitude \color{black} of the strain is given by $V_0/E_F$, where $V_0$ represents the modulation potential originating from the interfacial strain and $E_F$ is the Fermi energy. A value of $V_0=0.29\pm{0.001}\,\mathrm{meV}$ ($\approx$ 3\% of the Fermi energy) at $6\,\mathrm{K}$ is extracted for a $130\,\mathrm{nm}$ high periodic array, akin to the strain potentials induced by negative resists commonly used for creating periodic strain profiles \cite{kamburov2012ballistic}.

For the measured range of temperatures $6\,\mathrm{K}$ $\le T \le$ $16\,\mathrm{K}$, \color{blue}\color{black} there is only a slight change in $V_0$ \color{black} (see Supplementary Information \ref{density and temperature}), in agreement with the small variation in the thermal expansion at these temperatures. Also the 2DEG density dependence of $V_0$ over the accessible density range is small. In contrast, a strong dependence is observed for the value of $V_0$ on the height of the periodic array, indicating an increase of strain with metal thickness (\color{blue}\color{black} Supplementary \cref{fig:fig s7})\color{black}.\par

\begin{figure}[t]
    \includegraphics[width=0.5\textwidth]{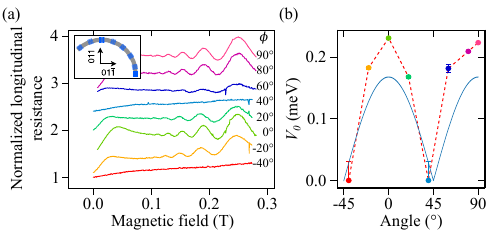}
\caption{Dependence of CO amplitude and modulation potential on the angle of orientation of the axis of periodic gates. (a) Oscillatory part of the normalized magnetoresistance at $T=10\,\mathrm{K}$ for several angles of orientation of the axis of the periodic gate.  $\phi = 0 \degree$ corresponds to the crystal direction [011]. The curves are shifted for clarity. Inset: Schematic layout of the Hallbar used for the measurements. The periodic gate was made up of $130\,\mathrm{nm}$ of Au on top of a continuous layer of Ti/Au $(3\,\mathrm{nm}/20\,\mathrm{nm})$ (b) Extracted modulation potential as a function of the orientation of the periodic gate (
coloured dots following the colours of the traces in \cref{fig:fig 3}(a) and dashed red  \color{black} lines as a guide for the eye).
Solid lines show the simulated values for a 2DEG $125\,\mathrm{nm}$ below the metal interface.}
\label{fig:fig 3}
\end{figure}

The strain and piezoelectric origin of the observed magnetoresistance oscillations can be further evidenced by examining the effect on the crystal orientation of the periodic gate.
To do this, we study the COs in a periodic gate device with different orientations of the axis of the periodic potential (mesa is kept aligned with the axis of the periodic potential). As shown in \cref{fig:fig 3}, the amplitude of COs depend strongly on the angle of orientation of the axis of the periodic potential, maximizing at an angle of 0$^{\circ}$ and 90$^{\circ}$ and minimizing at 40$^{\circ}$ and -40$^{\circ}$. It is consistent with the angle dependence of the piezoelectric effect in GaAs \cite{skuras1997anisotropic}, which is maximum along the [011] equivalent directions.  This evidences the role of the piezoelectricity in GaAs for the propagation of strain from the metallic gates to the 2DEG.
\color{black} The substantial value of modulation potential from a metallic periodic gate underscores the impact of strain from metallic electrostatic gates on the 2DEG. \color{red}  \color{black} \par

\color{blue}\color{black}

To gain further insight into how the strain induced by the metallic LSL propagates through the heterostructure and affects the buried 2DEG, finite-element simulations were performed using COMSOL Multiphysics.
The mechanical response of the GaAs substrate to the forces expected by the metallic LSL was calculated using the \emph{structural mechanics} module. A detailed description of the simulated geometry, derivation of the acting forces, and implementation is provided in the Supplementary Information \ref{numerical simulation}, \ref{model for forces} and \ref{influence of superlattice periodicity}.

Since our transport experiments were conducted at cryogenic temperatures around $5\,\mathrm{K}$, the full strain tensor was calculated for a temperature change of $300\,\mathrm{K}$. Fig. \ref{fig:fig 4}.(a) shows the calculated shear strain component $S_{xy}$ along a plane through the sample. The shear strain originates at the step edges of the metal stripes and propagates into the GaAs substrate, gradually relaxing with increasing distance from the surface. Within the first-order piezoelectric approximation, only the shear strain contributes to the piezoelectric polarization and, consequently, to the electrostatic potential modulation. The remaining components of the strain tensor are presented in the Supplementary Information \ref{model for forces} .
Using the piezoelectric coupling tensor of GaAs, the corresponding piezoelectric displacement field and electrostatic potential was calculated. The modulation amplitude $V_0$ was extracted  as a function of the crystal orientation, 2DEG depth ($D$) and LSL geometry (periodicity $P$ and bar height $H_{bar}$).

Fig. \ref{fig:fig 3}(b) compares the experimentally extracted modulation amplitudes $V_0$ with the values obtained from the numerical model for different crystal orientations. The solid blue line shows the calculated angular dependence of $V_0$ at the experimental depth \color{black} $D$. Experimental and numerical values are in good agreement for both modulation amplitude and crystal orientation. This agreement provides additional strong evidence that the observed commensurability oscillations originate from the strain field induced by the metallic superlattice.



The numerical simulation further allows us to systematically investigate the key parameters influencing the strain-induced potential modulation, in particular the dependence of $V_0$ on $D$, $H_{bar}$ and $P$ can be evaluated independently.


\color{blue}\color{black}Fig. \ref{fig:fig 4}(b) shows the calculated modulation amplitude $V_0$ as a function of $D$ for different values of $P$ (including the experimentally investigated case of $P=240\,\mathrm{nm}$). Starting from the surface 
of the GaAs crystal\color{black}, $V_0$ initially increases and reaches a maximum at $D \approx 50\,\mathrm{nm}$. At larger depths, the modulation amplitude decreases approximately exponentially, with a characteristic decay length of about $100\,\mathrm{nm}$. A change in $H_{bar}$ leads to an approximately linear response of $V_0$. The results of a detailed calculation can be found in the Supplementary Information fig. \ref{fig:fig s12}. The combined dependence of $V_0$ on $D$ and $H_{bar}$ is summarized in fig. \ref{fig:fig 4}(c), which can be used to estimate the maximum bar height required to limit the modulation amplitude at a given depth to a given value.

The tail of the strain penetration was found to be very sensitive to the LSL periodicity. Fig. \ref{fig:fig 4}(b) compares the calculated $D$ dependence of $V_0$ for three different values of  \color{red}\color{black}$P$\color{blue}\color{black}. Increasing the periodicity enhances the modulation amplitude and slightly modifies its depth dependence.
\color{blue}\color{black} The origin of this behavior can be understood from the shear strain distribution shown in fig. \ref{fig:fig 4}(a) and fig. \ref{fig:fig s12}.The strain field generated at the edges of neighboring metallic bars propagates into the GaAs substrate and gradually spreads laterally with increasing $D$. As a result, the strain fields from neighboring LSL bars overlap, reducing the spatial modulation of the strain and the corresponding piezoelectric potential. For larger $P$, this overlap and cancellation occurs at deeper depth, resulting in a stronger potential modulation at a given $D$. As we argue in the Discussion section, this effect is essential for understanding the mobility reduction observed under uniform metal gates.

\color{black}

\section{Discussion} \label{sec:Discussion}

\begin{figure}[htbp]
\centering
\includegraphics[width=0.5\textwidth]{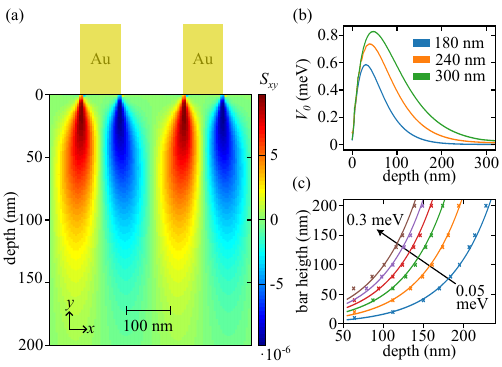}
\caption{{\color{blue}\color{black}(a) Heat map of the shear component $S_{xy}$ of the strain tensor at a temperature  $300\,\mathrm{K}$ below the deposition temperature 
of the sample\color{black}. The positions of the Au LSL bars are indicated by the gold colored rectangles above the map. (b) Simulated modulation amplitude $V_0$ as a function of 2DEG depth for different LSL periodicities (c) Minimum bar height required to obtain a given modulation amplitude $V_0$ at a specified 2DEG depth. Crosses represent the data extracted from the numerical simulation, while solid lines represent the empirical formula \color{red}\color{black}
\protect\\[0.3em]
\mbox{}\(\displaystyle H_{\mathrm{bar}}=\left(0.647\,\mathrm{meV}^{-1} \cdot P-103.93\,\mu\mathrm{m}\,\mathrm{eV}^{-1}\right)\cdot V_0 \, \cdot \)\hfill
\protect\\[-0.1em]
\mbox{}\hfill\(\displaystyle {}\exp\left(0.019\,\mathrm{nm}^{-1}\cdot D\right)\),
\protect\\[0.2em]
\color{red}\color{black}
with periodicity \(P\). The six curves correspond to modulation amplitudes ranging from \(0.05\,\mathrm{meV}\) to \(0.3\,\mathrm{meV}\).}}\color{black}
\label{fig:fig 4}
\end{figure}

\color{blue}\color{black} The observations of commensurability oscillations and associated COMSOL modelling clearly show that metallic Schottky gates cooled to cryogenic temperatures can induce strain in a GaAs based 2DEG.  As seen in \cref{fig:fig 4}, the strain is reduced with $D$ into the semiconductor. 
In fact, our simulations show that the penetration of the strain can be described by a monotonically increasing function of the thickness of the metal gate and the distance from the metal-semiconductor interface to the 2DEG. (See fig. \ref{fig:fig 4}(c)).  This, together with the continuous decrease in the 2DEG mobility with the increase in metal thickness, agrees with the idea that the additional scattering is a result of the strain induced by the gate.  Conversely, it is not consistent with other possible origins, such as inhomogeneity or pinned surface charge impurities in the metal layer and the screening of the electron-interactions in the 2DEG. \par
\FloatBarrier

We should now consider how the strain at the metal-semiconductor interface induces disorder 
in a normal gate with uniform thickness. 
A natural explanation would be the granular nature of the deposited film. The strain in a granular film is expected to be inhomogeneous due to the presence of the distribution of grain size and boundaries, with the strain likely being different at the boundaries and at the interior of the grains. The piezoelectric nature of GaAs will propagate the disordered strain distribution to the 2DEG, where it will result in a Fermi-level inhomogeneity which augments the disorder potential. As shown in our simulations, the strain propagates further into the GaAs for larger features (see \cref{fig:fig 4} and supplementary \cref{fig:fig s10}).
One would therefore expect that it is the largest grains that will be responsible for most of the mobility reduction 

\color{blue}\color{black}It is possible to test these ideas in a more quantitative way, using the mobility dependence on the gate metal thickness.  For this we need to make two further assumptions, first that the size distributions of the metal grains are similar in the in-plane and the growth direction. Second, we assume a self-similar scaling of the grain distribution \cite{Scaling_properties}, so that the mean size also will give information about the largest grains, since theses determine the random strain potential.  The grain size $L_{grain}$ in a thin metal film is expected to depend on the thickness $d$ as 
\begin{equation}
L_{grain} \propto d^{a},
\end{equation}
where $a$ is the so-called growth exponent. In the case of a three-dimensional solid-phase energy balance $a = 0.4$, which is also observed experimentally (see \cite{Grain-size} and references therein). 
Furthermore, since the strain potential $V_{0}$ is much smaller than the Fermi energy, the Born approximation should be valid, so that 
\begin{equation}
\mu_{strain} \propto V_{0}^{-2},
\end{equation}
with 
\begin{equation}
1/\mu = 1/\mu_{ug} + 1/\mu_{strain}, 
\end{equation}
$\mu_{ug}$ the measured mobility on an ungated sample, $\mu$ the measured mobility with a gate and for the same carrier density as the ungated sample, and $\mu_{strain}$ the strain-dependent part of the mobility. 
Since $V_0$ was found in our simulations to be proportional to the gate thickness (see the Supplementary Information \ref{influence of superlattice periodicity}) and in agreement with the experiments, \color{black}
one should therefore expect that 
\begin{equation}
\mu_{strain} \propto d^{-2a}.
\end{equation}
Using the data from \ref{sec:Results} to deduce $\mu_{strain}$ for $n_{s} = (8.5 \pm 0.2) \times 10^{14}\,\mathrm{m^{-2}} $, we find indeed a power law dependence on the gate thickness with an exponent $\approx 0.85$, in reasonable agreement with the above model. (See fig. \ref{fig:fig 1}(b) and supplementary  \cref{fig:fig s2}.)

\color{blue}

\color{blue}\color{black}
Our experimental results and simulations indicate three strategies for limiting the strain induced disorder. First, by burying the 2DEG  deeply into the semiconductor, and/or by using very thin gate metal, it is possible to reduce the strain sufficiently for its contribution to the disorder potential to be negligible. However, this typically requires a distance between the 2DEG and the metal/semiconductor interface that precludes the realisation of nanoscale gates such as quantum point contacts. (However, a recent experiment has overcome this limitation by using etch-defined quantum point contacts \cite{Alkalai2026e/4charge}). Furthermore, there is a limit to how thin a normal metal gate can be made (though it may be of interest to use a van der Waals conducting layer as a gate material).\par
A second method, incorporating a thin Atomic Layer Deposited (ALD) HfO\textsubscript{2} layer between the metallic gate layer and the GaAs heterostructure has been used successfully in a few studies \cite{dutta2022distinguishing, dutta2022isolated,veillon2024observation}. It is not necessarily the thermal properties of HfO\textsubscript{2} that plays the main role 
($\alpha$\textsubscript{HfO\textsubscript{2}} is very similar to $\alpha$\textsubscript{GaAs}, and HfO\textsubscript{2} is  stiffer than GaAs). However in addition to increasing 
the distance between \color{black} the 2DEG and the gate metal it is possible that the inserted layer also modifies the grain structure at the interface. As seen in supplementary \cref{fig:fig s1}, the gate layers need careful optimisation to minimise the disorder. We intend to explore this strategy further in future work.\par
Finally, our results also indicate that specially designed gates could be used to reduce the strain induced disorder.  For example, 
a possibility could be to employ a square LSL with a small lattice period, in order to limit the effective grain size in a similar fashion to what we demonstrated in our present experiments.  
\color{black}

\color{black}


\color{red}
\color{black}
\section*{Acknowledgements} \label{sec:acknowledgements}
The authors thank Frank Glas, Alexei Chepelianskii, Gwendal Fève, Anne Anthore and Frederic Pierre for fruitful discussions.  This work was supported by the French RENATECH network, the Agence National de la Recherche (ANR-20-CE30-0015-01 and ANR-22-PETQ-0012), the European Research Council (ERC-2020-SyG-951451), and DIM NANO-K (convention \#1763). 
\bibliography{biblio.bib}

@article{bartolomei2020collider,
  title={A collider for anyons},
  author={Bartolomei, H and Kumar, M and Bisognin, R and Marguerite, A and Berroir, JM and Bocquillon, B Pla{\c{c}}ais and Cavanna, A and Dong, Q and Gennser, U and Jin, Y and others},
  journal={Science},
  volume={368},
  pages={173--177},
  year={2020}
}

@article{nakamura2020direct,
  title={Direct observation of anyonic braiding statistics},
  author={Nakamura, James and Liang, Shuang and Gardner, Geoffrey C and Manfra, Michael J},
  journal={Nature Physics},
  volume={16},
  number={9},
  pages={931--936},
  year={2020},
  publisher={Nature Publishing Group UK London}
}

@article{glidic2023cross,
  title={Cross-correlation investigation of anyon statistics in the $\nu$= 1/3 and 2/5 fractional quantum Hall states},
  author={Glidic, P and Maillet, O and Aassime, A and Piquard, C and Cavanna, A and Gennser, Ulf and Jin, Y and Anthore, A and Pierre, F},
  journal={Physical Review X},
  volume={13},
  number={1},
  pages={011030},
  year={2023},
  publisher={APS}
}

@article{glidic2024signature,
  title={Signature of anyonic statistics in the integer quantum Hall regime},
  author={Glidic, P and Petkovic, I and Piquard, C and Aassime, A and Cavanna, A and Jin, Y and Gennser, U and Mora, C and Kovrizhin, D and Anthore, A and others},
  journal={Nature Communications},
  volume={15},
  number={1},
  pages={6578},
  year={2024},
  publisher={Nature Publishing Group UK London}
}

@article{veillon2024observation,
  title={Observation of the scaling dimension of fractional quantum Hall anyons},
  author={Veillon, A and Piquard, C and Glidic, P and Sato, Y and Aassime, A and Cavanna, A and Jin, Y and Gennser, U and Anthore, A and Pierre, F},
  journal={Nature},
  volume={632},
  pages={517},
  year={2024},
  publisher={Nature Publishing Group UK London}
}

@article{ruelle2023comparing,
  title={Comparing fractional quantum Hall Laughlin and Jain topological orders with the anyon collider},
  author={Ruelle, M{\'e}lanie and Frigerio, Elric and Berroir, J-M and Pla{\c{c}}ais, Bernard and Rech, J{\'e}r{\^o}me and Cavanna, Antonella and Gennser, Ulf and Jin, Y and F{\`e}ve, Gwendal},
  journal={Physical Review X},
  volume={13},
  number={1},
  pages={011031},
  year={2023},
  publisher={APS}
}

@article{lee2023partitioning,
  title={Partitioning of diluted anyons reveals their braiding statistics},
  author={Lee, June-Young M and Hong, Changki and Alkalay, Tomer and Schiller, Noam and Umansky, Vladimir and Heiblum, Moty and Oreg, Yuval and Sim, H-S},
  journal={Nature},
  volume={617},
  number={7960},
  pages={277--281},
  year={2023},
  publisher={Nature Publishing Group UK London}
}

@article{dutta2022distinguishing,
  title={Distinguishing between non-abelian topological orders in a quantum Hall system},
  author={Dutta, Bivas and Yang, Wenmin and Melcer, Ron and Kundu, Hemanta Kumar and Heiblum, Moty and Umansky, Vladimir and Oreg, Yuval and Stern, Ady and Mross, David},
  journal={Science},
  volume={375},
  number={6577},
  pages={193--197},
  year={2022},
  publisher={American Association for the Advancement of Science}
}

@article{kundu2023anyonic,
  title={Anyonic interference and braiding phase in a Mach-Zehnder interferometer},
  author={Kundu, Hemanta Kumar and Biswas, Sourav and Ofek, Nissim and Umansky, Vladimir and Heiblum, Moty},
  journal={Nature physics},
  volume={19},
  number={4},
  pages={515--521},
  year={2023},
  publisher={Nature Publishing Group UK London}
}

@article{dutta2022isolated,
  title={Isolated ballistic non-abelian interface channel},
  author={Dutta, Bivas and Umansky, Vladimir and Banerjee, Mitali and Heiblum, Moty},
  journal={Science},
  volume={377},
  number={6611},
  pages={1198--1201},
  year={2022},
  publisher={American Association for the Advancement of Science}
}

@article{weiss1989magnetoresistance,
  title={Magnetoresistance oscillations in a two-dimensional electron gas induced by a submicrometer periodic potential},
  author={Weiss, Dieter and Klitzing, KV and Ploog, K and Weimann, G},
  journal={Europhysics Letters},
  volume={8},
  number={2},
  pages={179},
  year={1989},
  publisher={IOP Publishing}
}

@article{winkler1989landau,
  title={Landau band conductivity in a two-dimensional electron system modulated by an artificial one-dimensional superlattice potential},
  author={Winkler, RW and Kotthaus, JP and Ploog, K},
  journal={Physical Review Letters},
  volume={62},
  number={10},
  pages={1177},
  year={1989},
  publisher={APS}
}

@article{beenakker1989guiding,
  title={Guiding-center-drift resonance in a periodically modulated two-dimensional electron gas},
  author={Beenakker, CWJ},
  journal={Physical review letters},
  volume={62},
  number={17},
  pages={2020},
  year={1989},
  publisher={APS}
}

@article{beton1990magnetoresistance,
  title={Magnetoresistance of a two-dimensional electron gas in a strong periodic potential},
  author={Beton, PH and Alves, ES and Main, PC and Eaves, L and Dellow, MW and Henini, M and Hughes, OH and Beaumont, SP and Wilkinson, CDW},
  journal={Physical Review B},
  volume={42},
  number={14},
  pages={9229},
  year={1990},
  publisher={APS}
}

@article{peeters1992electrical,
  title={Electrical and thermal properties of a two-dimensional electron gas in a one-dimensional periodic potential},
  author={Peeters, FM and Vasilopoulos, P},
  journal={Physical Review B},
  volume={46},
  number={8},
  pages={4667},
  year={1992},
  publisher={APS}
}

@article{mirlin1998weiss,
  title={Weiss oscillations in the presence of small-angle impurity scattering},
  author={Mirlin, AD and W{\"o}lfle, P},
  journal={Physical Review B},
  volume={58},
  number={19},
  pages={12986},
  year={1998},
  publisher={APS}
}

@article{endo2000envelope,
  title={Envelope of commensurability magnetoresistance oscillation in unidirectional lateral superlattices},
  author={Endo, A and Katsumoto, S and Iye, Yasuhiro},
  journal={Physical Review B},
  volume={62},
  number={24},
  pages={16761},
  year={2000},
  publisher={APS}
}

@article{kamburov2012ballistic,
  title={Ballistic transport of (001) GaAs two-dimensional holes through a strain-induced lateral superlattice},
  author={Kamburov, D and Shapourian, H and Shayegan, M and Pfeiffer, LN and West, KW and Baldwin, KW and Winkler, R},
  journal={Physical Review B—Condensed Matter and Materials Physics},
  volume={85},
  number={12},
  pages={121305},
  year={2012},
  publisher={APS}
}

@article{willett1987observation,
  title={Observation of an even-denominator quantum number in the fractional quantum Hall effect},
  author={Willett, Robert and Eisenstein, James P and St{\"o}rmer, Horst L and Tsui, Daniel C and Gossard, Arthur C and English, JH},
  journal={Physical review letters},
  volume={59},
  number={15},
  pages={1776},
  year={1987},
  publisher={APS}
}

@article{wang2022even,
  title={Even-denominator fractional quantum Hall state at filling factor $\nu$= 3/4},
  author={Wang, Chengyu and Gupta, Adbhut and Singh, Siddharth Kumar and Chung, YJ and Pfeiffer, LN and West, KW and Baldwin, KW and Winkler, Roland and Shayegan, Mansour},
  journal={Physical Review Letters},
  volume={129},
  number={15},
  pages={156801},
  year={2022},
  publisher={APS}
}

@article{eisenstein1992new,
  title={New fractional quantum Hall state in double-layer two-dimensional electron systems},
  author={Eisenstein, JP and Boebinger, GS and Pfeiffer, LN and West, KW and He, Song},
  journal={Physical review letters},
  volume={68},
  number={9},
  pages={1383},
  year={1992},
  publisher={APS}
}

@article{suen1992observation,
  title={Observation of a $\nu$= 1/2 fractional quantum Hall state in a double-layer electron system},
  author={Suen, YW and Engel, LW and Santos, MB and Shayegan, M and Tsui, DC},
  journal={Physical review letters},
  volume={68},
  number={9},
  pages={1379},
  year={1992},
  publisher={APS}
}

@article{luhman2008observation,
  title={Observation of a fractional quantum Hall state at $\nu$= 1/4 in a wide GaAs quantum well},
  author={Luhman, DR and Pan, W and Tsui, DC and Pfeiffer, LN and Baldwin, KW and West, KW},
  journal={Physical Review Letters},
  volume={101},
  number={26},
  pages={266804},
  year={2008},
  publisher={APS}
}

@article{wang2023next,
  title={Next-generation even-denominator fractional quantum Hall states of interacting composite fermions},
  author={Wang, Chengyu and Gupta, Adbhut and Madathil, Pranav T and Singh, Siddharth K and Chung, Yoon Jang and Pfeiffer, Loren N and Baldwin, Kirk W and Shayegan, Mansour},
  journal={Proceedings of the National Academy of Sciences},
  volume={120},
  number={52},
  pages={e2314212120},
  year={2023},
  publisher={National Acad Sciences}
}

@article{huang2024evidence,
  title={Evidence for Topological Protection Derived from Six-Flux Composite Fermions},
  author={Huang, Haoyun and Hussain, Waseem and Myers, SA and Pfeiffer, LN and West, KW and Baldwin, KW and Cs{\'a}thy, GA},
  journal={Nature Communications},
  volume={15},
  number={1},
  pages={1461},
  year={2024},
  publisher={Nature Publishing Group UK London}
}

@article{chung2021ultra,
  title={Ultra-high-quality two-dimensional electron systems},
  author={Chung, Yoon Jang and Villegas Rosales, KA and Baldwin, KW and Madathil, PT and West, KW and Shayegan, M and Pfeiffer, LN},
  journal={Nature Materials},
  volume={20},
  number={5},
  pages={632--637},
  year={2021},
  publisher={Nature Publishing Group UK London}
}

@article{bocquillon2014electron,
  title={Electron quantum optics in ballistic chiral conductors},
  author={Bocquillon, Erwann and Freulon, Vincent and Parmentier, Fran{\c{c}}ois D and Berroir, Jean-Marc and Pla{\c{c}}ais, Bernard and Wahl, Claire and Rech, J{\'e}rome and Jonckheere, Thibaut and Martin, Thierry and Grenier, Charles and others},
  journal={Annalen der Physik},
  volume={526},
  number={1-2},
  pages={1--30},
  year={2014},
  publisher={Wiley Online Library}
}

@article{ruelle2024time-resolved,
  title={Time-resolved sensing of electromagnetic fields with single-electron interferometry},
  author={Ruelle, M{\'e}lanie and Frigerio, Elric and  Baudin, Emmanuel and  Berroir, Jean-Marc and Placais, Bernard and Gr{\'e}maud, Beno{\^\i}t and Jonckheere, Thibaut and Martin, Thierry and Rech, J{\'e}r{\^o}me and Cavanna, Antonella and Gennser, Ulf and Jin, Yong and M{\'e}nard, Gerbold and F{\`e}ve, Gwendal},
  journal={Science},
  volume={389},
  number={6755},
  pages={7695},
  year={2025},
}

@article{schmiedeshoff2006versatile,
  title={Versatile and compact capacitive dilatometer},
  author={Schmiedeshoff, GM and Lounsbury, AW and Luna, DJ and Tracy, SJ and Schramm, AJ and Tozer, SW and Correa, VF and Hannahs, ST and Murphy, TP and Palm, EC and others},
  journal={Review of Scientific Instruments},
  volume={77},
  number={12},
  year={2006},
  publisher={AIP Publishing}
}

@article{pamato2018thermal,
  title={The thermal expansion of gold: point defect concentrations and pre-melting in a face-centred cubic metal},
  author={Pamato, Martha G and Wood, Ian G and Dobson, David P and Hunt, Simon A and Vo{\v{c}}adlo, Lidunka},
  journal={Journal of Applied Crystallography},
  volume={51},
  number={2},
  pages={470--480},
  year={2018},
  publisher={International Union of Crystallography}
}

@article{loureno2004thermal,
  title={Thermal expansion contribution to the temperature dependence of excitonic transitions in GaAs and AlGaAs},
  author={Loureno, SA and Dias, IFL and Duarte, Jos{\'e} Leonil and Laureto, Edson and Po{\c{c}}as, LC and Toginho Filho, DO and Leite, JR},
  journal={Brazilian Journal of Physics},
  volume={34},
  pages={517--525},
  year={2004},
  publisher={SciELO Brasil}
}

@article{skuras1997anisotropic,
  title={Anisotropic piezoelectric effect in lateral surface superlattices},
  author={Skuras, E and Long, AR and Larkin, IA and Davies, JH and Holland, MC},
  journal={Applied Physics Letters},
  volume={70},
  pages={871--873},
  year={1997}
}

@article{singh2024topological,
  title={Topological phase transition between Jain states and daughter states of the $\nu$= 1/2 fractional quantum Hall state},
  author={Singh, SK and Wang, C and Tai, CT and Calhoun, CS and Villegas Rosales, KA and Madathil, PT and Gupta, A and Baldwin, KW and Pfeiffer, LN and Shayegan, M},
  journal={Nature Physics},
  volume={20},
  number={8},
  pages={1247--1252},
  year={2024},
  publisher={Nature Publishing Group UK London}
}

@article{Alkalai2026e/4charge,
    title={Observation of e/4 charge at $\nu =1/2$ in GaAs},
    author={Alkalai, T and Hajigeorgiou, E and Gupta, A and Senapati, T and Tiwari, P and Tai, C-T and Singh, SK and Baldwin, KW and Pfeiffer, LN and Shayegan, M and Banerjee, M and Heiblum, M}, 
    journal={arXiv:2602.08468}, 
    year={2026}
}

@article{Scaling_properties,
    title={Scaling properties of ultrasonic attenuation and grain size in metals},
    author={Nicoletti, DW},
    journal={Drexel University ProQuest Dissertations \& Theses:1991.9128534},
    year={1991}
}

@article{Grain-size,
    title={On the grain size-thickness correlation for thin films},
    author = {Dulmaa, A and Cougnon, FG and Dedoncker, R and Depla, D},
    journal = {Acta Materialia},
    volume = {212}, 
    pages ={116896},
    year = {2021}
}

@article{davies1994theory,
  title={Theory of potential modulation in lateral surface superlattices},
  author={J. H. Davies and I. A. Larkin},
  journal={Physical Review B},
  volume={49},
  number={7},
  pages={4800},
  year={1994},
  publisher={APS}
}

@article{Larkin1997,
  title={Theory of potential modulation in lateral surface superlattices. II. Piezoelectric effect},
  author={Larkin, Ivan A and Davies, John H and Long, Andrew R and Cusc{\'o}, Ramon},
  journal={Physical Review B},
  volume={56},
  number={23},
  pages={15242},
  year={1997},
  publisher={APS}
}

@article{hu1991stress,
  title={Stress-related problems in silicon technology},
  author={Hu, SM},
  journal={Journal of applied physics},
  volume={70},
  number={6},
  pages={R53--R80},
  year={1991},
  publisher={American Institute of Physics}
}

@article{murray2006mechanics,
  title={Mechanics of edge effects in anisotropic thin film/ substrate systems},
  author={Murray, Conal E},
  journal={Journal of Applied Physics},
  volume={100},
  number={10},
  year={2006},
  publisher={AIP Publishing}
}

\newpage
\setcounter{figure}{0}
\clearpage
\section{Supplementary Information} \label{Supplementary}

\setcounter{figure}{0}
\renewcommand{\thefigure}{S\arabic{figure}}
\renewcommand{\theHfigure}{S\arabic{figure}}
\subsection{Comparison of transport measurements for gated and ungated Hallbar devices}\label{comparison}
To observe the detrimental effects of the presence of a gate on the FQH states, we measured the longitudinal resistance of a gated device along with an adjacent ungated reference device at $T=10\,\mathrm{mK}$. Despite the presence of a $20\,\mathrm{nm}$ thick HfO\textsubscript{2} layer, grown via Atomic Layer Deposition (ALD), between the metal gate of $90\,\mathrm{nm}$ of Al and the sample surface,  as shown in supplementary \cref{fig:fig s1}, the FQH states are heavily suppressed. 

\begin{figure}[htbp]
    \centering
\includegraphics[width=0.5\textwidth]{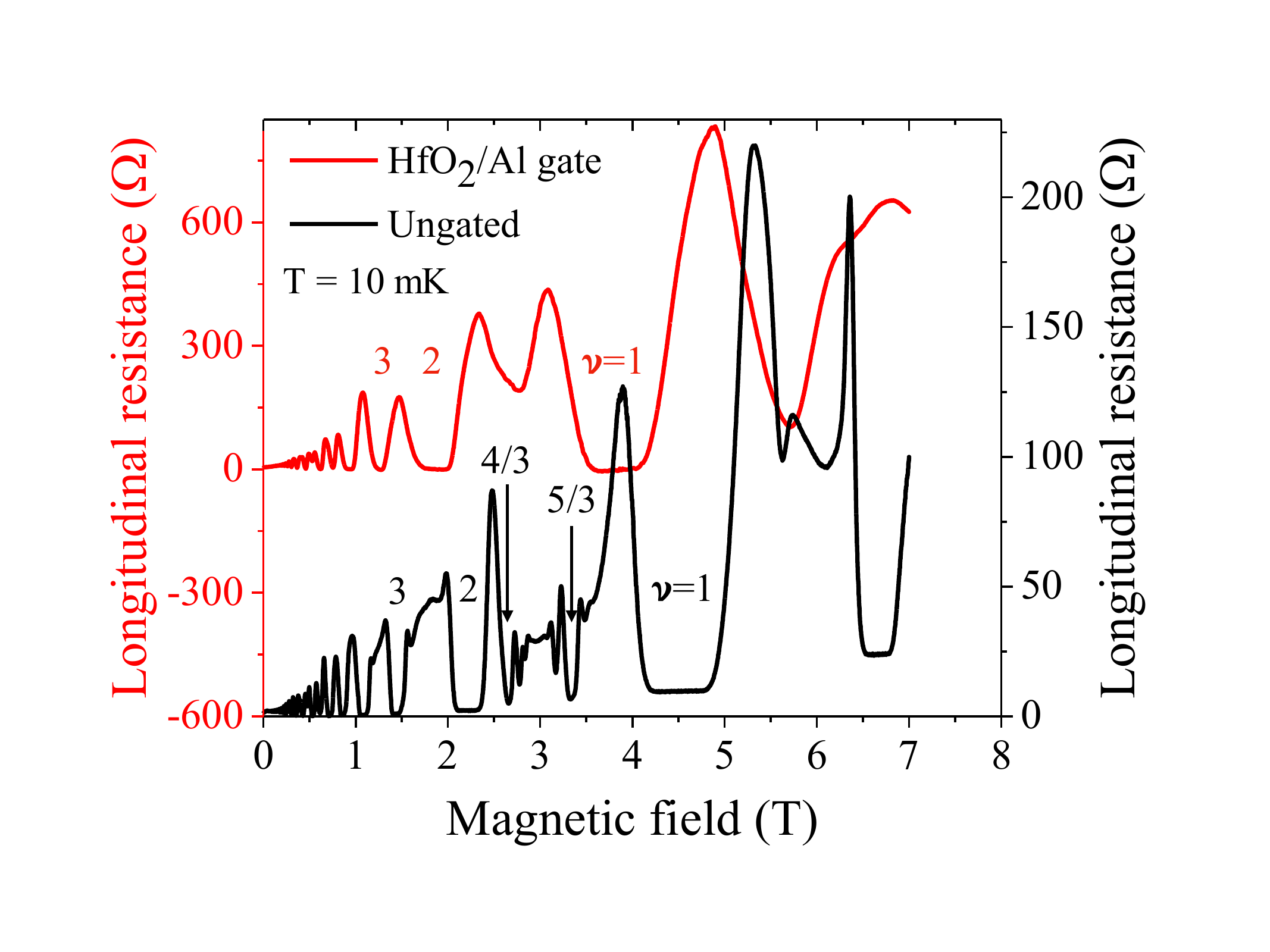}
\caption{Comparison of longitudinal resistance between a thin HfO\textsubscript{2} $(20\,\mathrm{nm})$ insulated gate with thick metallic cap (Al $90\,\mathrm{nm}$) device (red) and the reference ungated section (black). 
Filling factors $\nu$ of a few of the integer and fractional quantum Hall states are indicated. \color{black} The detrimental effect on the FQH states is evident in the red curve despite the presence of a HfO\textsubscript{2} layer. \color{blue}\color{black} The ungated section has a density $n_{s} = (1.5 \pm 0.2) \times 10^{15}\,\mathrm{m^{-2}} $ and a $6\,\mathrm{K}$ mobility $\mu = 1.23 \times 10^{2}\,\mathrm{m^{2} V^{-1} s^{-1}} $. \color{black}}
\label{fig:fig s1}
\end{figure}
\FloatBarrier

\subsection{Normalization of density-mobility data for unpatterned Al gated devices}\label{normalization}
\begin{figure}[htbp]
    \centering
    \includegraphics[width=0.5\textwidth]{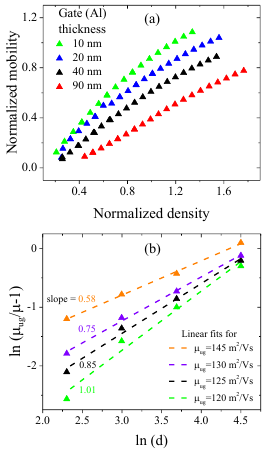}
\caption{(a) Normalized density-normalized mobility curves for the metallic gated devices corresponding to \cref{fig:fig 1}. The density and mobility of the gated devices are normalized to those of the ungated section of the same Hallbar device.
(b) $ln(\frac{\mu_{ug}}{\mu} - 1)$ vs.  $ln(d)$ for various values of $\mu_{ug}$ at the normalized density = 1. Note that $\mu_{ug} = 125 \ m^2/Vs$ is the average mobility measured for the ungated 2DEG, and $\mu_{ug}$ must be greater than the measured mobility.}
\label{fig:fig s2}
\end{figure}

The measurements shown in \cref{fig:fig 1}b of the main text represent the untreated data obtained on different samples from the same wafer. Supplementary \cref{fig:fig s2} shows the corresponding normalized data, where the density and the mobility of the gated Hallbar for each sample has been normalized to that of the ungated reference device on the same Hallbar adjacent to the gated devices.

\FloatBarrier

\subsection{Additional measurements on the periodic gate device shown in the main text}\label{additional measurements}
In order to visually distinguish the COs from the SdH oscillations, the periodic gate sample as shown in \cref{fig:fig 2} of the main text was measured at $T=12\,\mathrm{mK}$ and at relatively higher magnetic fields. As shown in the supplementary \ref{fig:fig s3}a, the SdH oscillations with a different frequency are clearly superimposed on the COs at higher magnetic fields. \par
Lithographically, each metal stripe is $90\,\mathrm{nm}$ in width and the array has a period of $241\,\mathrm{nm}$ (supplementary \ref{fig:fig s3}d). The green measurement scale obtained from the Scanning Electron Microscope corresponds to 31 periods and one additional metal stripe. Assuming the actual width of the one stripe to be exactly $90\,\mathrm{nm}$, the period of the array is estimated, from microscopy images, to be $242.8\,\mathrm{nm}$.

\begin{figure}[htbp]
    \centering
    \includegraphics[width=0.5\textwidth]{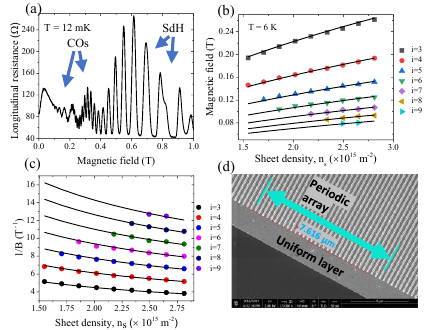} 
\caption{(a) shows the longitudinal resistance of the periodic gate device of \cref{fig:fig 2} of the main text,  at $12\,\mathrm{mK}$, showing both COs at lower and SdH oscillations superimposed on COs at higher magnetic fields, respectively. (b) and (c) show the variation of the magnetic field (b) and its inverse (c) at the longitudinal resistance minima for several densities, following the commensurability condition (eq. 1). \color{blue}\color{black} The data points represent the CO minima for the corresponding index i and the solid lines represent eq. 1. A good agreement between the two is visually evident. \color{black} (d) shows an electron microscopy image of the gate where the uniform layer is $20\,\mathrm{nm}$ of Al and the periodically patterned array is $130\,\mathrm{nm}$ of Al in height. } 
\label{fig:fig s3}
\end{figure}
\FloatBarrier

\subsection{Fitting procedure for Commensurability oscillations}\label{fitting procedure}
To extract the oscillatory part, the peaks and the troughs of the normalized magnetoresistance were locally fitted with polynomials of order 7 near the extrema. The maximum and minimum values thus were spline-connected to create the CO envelope, and the average of them provided the background magnetoresistance. This was subtracted from the data to obtain the oscillatory part.\cite{kamburov2012ballistic}
To fit the oscillatory part with eq. 2, standard Levenberg-Marquardt algorithm was used with $\tau_{CO}$, $V_0$ and $n$ as free parameters. The period of the lateral superlattice (metallic array)  was fixed at $241\,\mathrm{nm}$. The mobility was kept fixed at the value obtained from density-mobility measurements at respective temperature and density. 

\begin{figure}[htbp]
    \centering
    \includegraphics[width=0.5\textwidth]{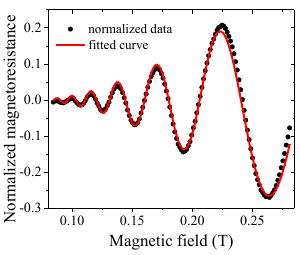}
\caption{Commensurability oscillations in the form of the fractional change of the magnetoresistance normalized to the zero-field resistance, $(\rho-\rho_0)/\rho_0$; where $\rho_0$ is the longitudinal resistance measured in the absence of magnetic field, as observed after background subtraction, in a sample with periodically patterned gate, measured at $6\,\mathrm{K}$ (black dots) at a density $n_{s} = 2.811 \times 10^{15}\,\mathrm{m^{-2}} $ and a mobility $\mu = 1.4535 \times 10^{2}\,\mathrm{m^{2} V^{-1} s^{-1}} $. 
The bar height of the periodic layer, made up of Al, was $130\,\mathrm{nm}$. The red curve shows the fitted curve as per eq. 2 in the main text. The value of $V_0=0.29\,\mathrm{meV}$ and $\tau=3.1\,\mathrm{ps}$ were obtained from the above fitting.}
\label{fig:fig s4}
\end{figure}
\FloatBarrier

\subsection{Density and temperature dependence of CO amplitude}\label{density and temperature}
Values of $V_0$ show a small variation with temperature, with $V_0=0.29 \pm{0.001}\,\mathrm{meV}$ at $T=6\,\mathrm{K}$ to $V_0=0.34 \pm{0.002}\,\mathrm{meV}$ at $T=16\,\mathrm{K}$. These values are obtained by fixing $\tau=3.16\,\mathrm{ps}$ and fitting the background subtracted normalized magnetoresistance for $V_0$ only. The $\tau=3.16\,\mathrm{ps}$ value is apriori obtained by fitting the $T=6$ K data with both $\tau$ and $V_0$ as open parameters. 
When changing the density 
$n_s$ over the accessible range, 
$V_0$ exhibits a variation from 
$V_0 = 0.22 \pm 0.003$ meV 
(for $n_s = 2.76 \times 10^{15} \, \mathrm{m}^{-2}$) to $V_0 = 0.24 \pm 0.001$ meV 
(for $n_s = 2.02 \times 10^{15} \, \mathrm{m}^{-2}$).
Density variations of the commensurability oscillations have been studied in details in ref. \cite{kamburov2012ballistic}.\color{black}

\begin{figure}[htbp]
    \centering
    \includegraphics[width=0.5\textwidth]{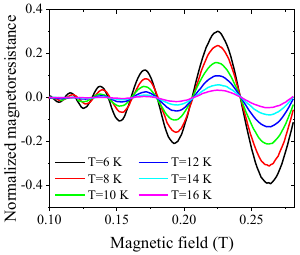}
\caption{Background subtracted normalized magnetoresistance data, $(\rho-\rho_0)/\rho_0$, at different temperatures, corresponding to \cref{fig:fig 2}d of the main text.}
\label{fig:fig s5}
\end{figure}
\FloatBarrier

\subsection{Commensurability oscillations in gated devices of different metals}\label{commensurability oscillations}

To verify the generality of the observed COs with respect to the type of metals used for the gate fabrication, we studied devices with periodically patterned gates of Ti/Au as described in the second section of the main text. These devices also resulted in pronounced COs and one such measurement is shown in supplementary \cref{fig:fig s6}. In this figure, the COs were obtained at $T=7\,\mathrm{K}$ for a range of densities, represented by the applied gate voltages. Each measurement was done at a fixed density. The curves represent the untreated data. As the gate voltage is reduced, the density and hence the 2DEG mobility is reduced and $R_{xx}$ shifts upwards in the plot.

\begin{figure}[htbp]
    \centering
    \includegraphics[width=0.5\textwidth]{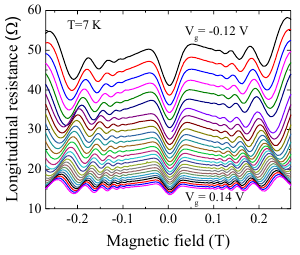}
\caption{Commensurability oscillations, at $T=7\,\mathrm{K}$, of devices with periodic gates of Ti/Au $(3\,\mathrm{nm})/(20\,\mathrm{nm})$ as the uniform layer and $130\,\mathrm{nm}$ of Au for the periodic layer, for different fixed gate voltages. The fixed gate voltage applied to the uniform gate spans from $-0.12\,\mathrm{V}$ (topmost curve), corresponding to a measured sheet density of $1.81\times \ 10^{15}\,\mathrm{m^{-2}}$, to $0.14\,\mathrm{V}$ (bottom curve), corresponding to a mesured density of $3.2\times 10^{15}\,\mathrm{m^{-2}}$ in steps of $0.01\,\mathrm{V}$. The plots in the figure represent the raw data.}
\label{fig:fig s6}
\end{figure}
\FloatBarrier

\subsection{Dependence of modulation potential on periodic gate height}\label{dependence on modulation potential}
Following the observation of the reduction in mobility with the increase in metal thickness in uniformly gated devices, indicating the increase in strain-originated disorder, it is imperative to quantify the dependence of strain on the metal thickness. Therefore, we studied two periodically gated devices with the same thickness of the uniform layer but different heights/thicknesses of the periodically patterned layer. The COs obtained from such devices are shown in supplementary \cref{fig:fig s7}.\par

First, the data for the $130\,\mathrm{nm}$ height sample (less noisy data) was fitted with eq. 2 of the main text as per the procedure explained in supplementary section A, with both $V_0$ and $\tau$ as the free parameters. A value of $V_0=0.301 \pm 0.006\,\mathrm{meV}$ and $\tau=4.45\pm0.3\,\mathrm{ps}$ were obtained. Then the data for the $60\,\mathrm{nm}$ sample was fitted for $V_0$ with the value of $\tau$ fixed at $\tau=4.45\,\mathrm{ps}$. The value of  $V_0=0.139 \pm 0.003\,\mathrm{meV}$ was obtained for the $60\,\mathrm{nm}$ sample. The ratio between the two $V_0$ values is very close to that between the two heights of the periodic layers, \color{blue}\color{black}  agreeing with the COMSOL Multiphysics modelling of the strain experienced by the 2DEG with the gate metal thickness (see section \ref{sec:Results}).\color{black} 

\begin{figure}[htbp]
    \centering
    \includegraphics[width=0.5\textwidth]{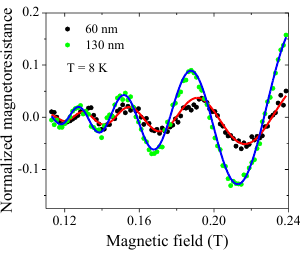}
\caption{Periodic layer height dependence of commensurability oscillations. Comparison of COs for different heights of the periodic layer with the vertical axis showing the fractional change of the magnetoresistance normalized to the zero-field resistance. Black and green data points show the COs in devices at $T=8\,\mathrm{K}$ with the the periodic layer (Au) height of $130\,\mathrm{nm}$ and $60\,\mathrm{nm}$, respectively. Red and blue lines are the corresponding fitting to the data obtained according to the procedure explained above. Both devices were fabricated on the same wafer and the uniform layer of Ti/Au $(3\,\mathrm{nm}/20\,\mathrm{nm})$ for both the devices were fabricated in the same metal deposition step. }
\label{fig:fig s7}
\end{figure}
\FloatBarrier

\subsection{Mesa orientation-dependence of longitudinal resistance}\label{mesa orientation}
In order to detect any possible correlation of the strain propagation (tuned by the angle of orientation of the mesa) to the mobility, we implemented the idea described in \cref{fig:fig 3} of the main text but with uniform unpatterned gates. Gated devices with different angles of mesa orientation were fabricated on the same Hallbar and the longitudinal resistance measured for all the device at a fixed density (controlled by a fixed potential to the metallic gate) are plotted in supplementary \cref{fig:fig s8}. The $R_{xx}$ values were normalized to the value obtained at a mesa angle of $40^{\circ}$. The uniform gate was made up of $300\,\mathrm{nm}$ of Al. \par

The longitudinal resistivity of the uniform unpatterned gated devices seems to follow, qualitatively, a similar dependence on the orientation angle of mesa as the estimated modulation potential in periodic gate devices. This strongly hints at a correlation between the strain from the presence of metallic gates and the consequent increase of disorder in the 2DEG due to the strain propagation. It should be noted that the magnitude of $R_{xx}$ dependence on the orientation angle is dependent on the 2DEG density; the effect is maximized at lower densities. \color{blue}\color{black}The peak at higher densities seems to occur at $\approx 20^{\circ}$ orientation angle, which may be attributed to measurement errors and normalization of the data at 40$^{\circ}$.\color{black}

\begin{figure}[htbp]
    \centering
    \includegraphics[width=0.5\textwidth]{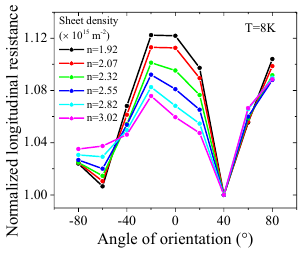}
\caption{Dependence of longitudinal resistance on the angle of orientation of the axis of the mesa. Longitudinal resistance for devices with different angles of orientation of the Hallbar axis, normalized to the resistance value for the $40^{\circ}$ device, measured at $8\,\mathrm{K}$ for several fixed sheet densities.}
\label{fig:fig s8}
\end{figure}
\FloatBarrier

\color{blue}\color{black}
\subsection{Sample used for measurements on patterned gates}\label{samples used} 
In supplementary \cref{fig:fig sPG} we show the cross-sectional view and the angle disposition of the sample used for the LSL gates. \color{black}
\begin{figure}[htbp]
    \centering
    \includegraphics[width=0.5\textwidth]{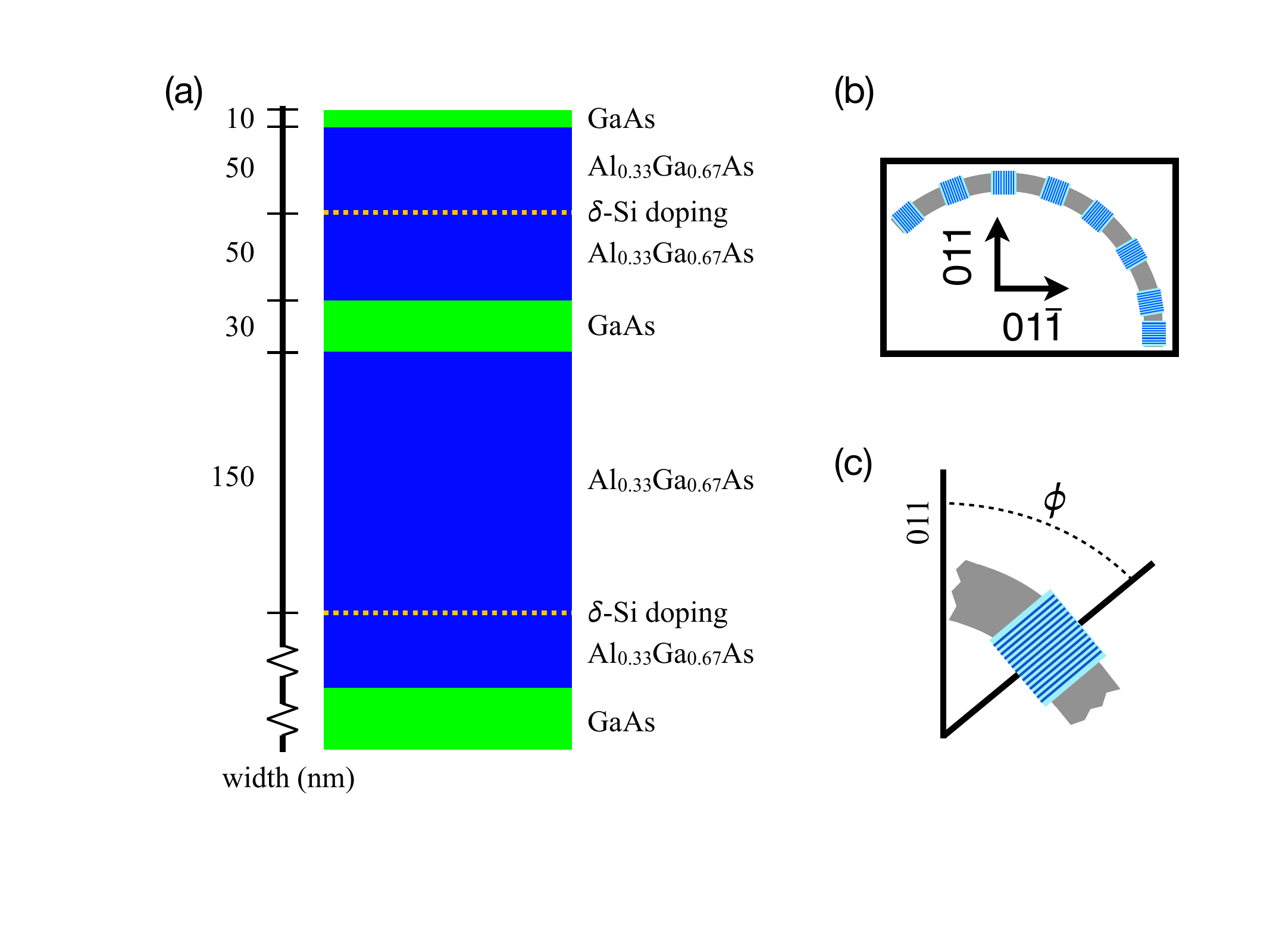}
\caption{\color{blue}\color{black} (a) Cross-sectional view of the sample on which the patterned gates were fabricated on. The center of the quantum well has a depth of $125\,\mathrm{nm}$. (b) Schematic view of the angle disposition  of the patterned gates and (c) of the angle $\phi$ between the lateral superlattice and the crystal direction [011] (bottom). \color{blue}} 
\label{fig:fig sPG}
\end{figure}

\color{blue}\color{black}

\subsection{Numerical simulation of strain propagation}\label{numerical simulation}
The strain produced by the patterned Al gate on the GaAs wafer was modeled using the \emph{Solid Mechanics} module of COMSOL Multiphysics (version 6.2). The symmetry of the LSL geometry and of the elasticity tensor of GaAs allows us to investigate the strain propagation in a two dimensional plane perpendicular to the lattice orientation. Additional structural features, such as the etched Hall bar, were neglected. For numerical stability, the layered heterostructure used in the experiments, shown in fig \ref{fig:fig sPG} was approximated by homogeneous GaAs. Likewise the continuous Au  film underneath the LSL bars was also replaced by homogeneous GaAs. Under the assumption of a homogeneous grainfree film, uniform thermal contraction of the of the continuous layer creates in plane stress and does not generate shear strain components. The continuous Au layer therefore primarily transmits the shear stress generated by the bars.


The calculation was performed on a two-dimensional $4\,\mathrm{\mu m}\times 4\,\mathrm{\mu m}$ square piece of GaAs. \color{black} The mesh configuration was a user-controlled triangular grid with a maximum element size of $8\,\mathrm{nm}$ and a minimum element size of $5\,\mathrm{nm}$. The growth rate was set to 1.1 with a curvature factor of 0.2 and the resulting resolution in the narrow regions was 1.0.                   
During the experimental deposition, the substrate temperature was kept at approximately $300\,\mathrm{K}$ by water cooling. At this temperature, the Au layer and the GaAs substrate were assumed to be free of thermal strain. 
By cooling the sample, mechanical strain develops due to the different thermal contractions of the two materials. Since the thermal expansion coefficients vary over the investigated temperature range, the effective values, averaged between $1\,\mathrm{K}$ to $300\,\mathrm{K}$ were chosen, with $\alpha_{\mathrm{Au}}=9.1\times 10^{-6}\,\mathrm{K^{-1}}$ for Au and $\alpha_{\mathrm{GaAs}}=2.5\times 10^{-6}\,\mathrm{K^{-1}}$ for GaAs.

Directly including the metallic gate in the finite element model leads to stress singularities at sharp material corners of the material interface. To avoid these numerical artifacts, the gate was not explicitly modeled. Instead, the mechanical forces produced by the Au bars were determined analytically with the model described in section \ref{sec: SpringModel} and applied as boundary loads on the GaAs block. The finite element simulation was then used to calculate the resulting displacement and strain fields within the substrate.\\

\subsection{Model for forces on the GaAs substrate} \label{sec: SpringModel}\label{model for forces}
\begin{figure}[htbp]
    \centering
    \includegraphics[width=0.5\textwidth]{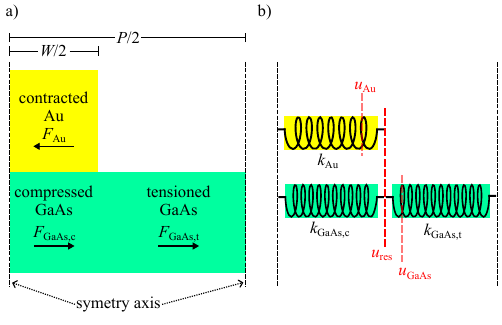}
\caption{\color{blue}\color{black}(a) Schematic illustration of the Au stripe and GaAs substrate and the resulting forces acting upon cooling. (b) Corresponding three spring model used to estimate the forces acting on the GaAs in the finite element simulation. The individual equilibrium positions as well as the combined equilibrium position after cooling are indicated.\\ \color{blue}}
\label{fig:fig s10}
\end{figure}

Figure \ref{fig:fig s10} illustrates the simplified model used to estimate the forces expected to act on the finite GaAs substrate. The Au bar, the GaAs directly underneath the bar, and the uncovered GaAs region within half a LSL period are represented by three elastically coupled springs. The springs are fixed on the symmetry axis of the sample. The spring constants are denoted by $k_{\mathrm{Au}}$, $k_{\mathrm{GaAs,c}}$, and $k_{\mathrm{GaAs,t}}$, respectively, where the subscripts $c$ and $t$ refer to the compressive and tensile regions of the substrate.\\
At the deposition temperature of $300\,\mathrm{K}$ the system is assumed to be fully relaxed, the equilibrium positions of all spring align. By cooling the different materials contract differently. The individual equilibrium positions of the springs shift according to the thermal contraction of the materials. All regions remain mechanically connected, thus they share a final equilibrium displacement position $u_{\mathrm{res}}$. This mechanical equilibrium requires the total force acting on the coupled system to vanish. The resulting displacement is therefore given by

\begin{equation*}
u_{\mathrm{res}}=\frac{k_{\mathrm{Au}}u_{\mathrm{Au}}+k_{\mathrm{GaAs,c}}u_{\mathrm{GaAs}}+k_{\mathrm{GaAs,t}}u_{\mathrm{GaAs}}}{k_{\mathrm{Au}}+k_{\mathrm{GaAs,c}}+k_{\mathrm{GaAs,t}}}.
\end{equation*}

Here $u_{\mathrm{Au}}=W/2 \cdot (1-\alpha_{\mathrm{Au}}\Delta T)$ represent the equilibrium width of the  half Au stripe after the thermal contraction in absence of mechanical constraints\color{red}\color{black}, with $W$ the width of a Au stripe at $300\,\mathrm{K}$. \color{blue}\color{black} Analogous expressions are used for the GaAs regions. The spring constants are given by $k=E\frac{H}{D_{mat}}$ where $E$ is the Young's modulus and $H$ and $D_{mat}$ are the height and width of the respective region. 
For the Au stripe, the spring constant follows directly from its geometrical dimensions.
For the GaAs substrate, an effective spring constant has to be determined because the spring model represents the finite GaAs domain used in the finite-element calculation. We therefore evaluate the displacement and forces as a function of the assumed GaAs thickness. With increasing substrate thickness $u_{\mathrm{res}}-u_{\mathrm{GaAs}}$ approaches zero, while the force excerted by the Au stripe approaches a limiting value. Beyond a thickness of several micrometers, both quantities depend only weakly on the assumed substrate thickness. We therefore use an effective GaAs thickness of $4\,\mathrm{\mu m}$, for which the model is already close to this limiting regime. This limit corresponds to an effectively fully constrained metallic gate, similar to the approximation used in analytic treats of strain induced by surface gates. \cite{davies1994theory, Larkin1997} \color{black} 

The spring model yields the effective force acting on the GaAs substrate along the $x$ direction, reading
\begin{equation*}
F_{x} = -k_{\mathrm{Au}} \cdot ( u_{\mathrm{res}} -u_{\mathrm{Au}}).
\end{equation*}

\begin{figure}[htbp]
    \centering
    \includegraphics[width=0.5\textwidth]{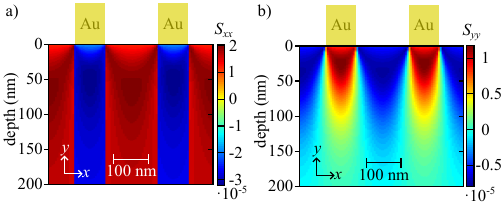}
\caption{\color{blue}\color{black}Heat maps of the strain components $S_{xx}$ (a) and $S_{yy}$ (b) of the calculated strain tensor. The corresponding shear strain component $S_{xy}$ is presented in Fig. \ref{fig:fig 4} (a) of the main text.\color{black}}
\label{fig:fig s11}
\end{figure}

The resulting strain tensor in the GaAs substrate was subsequently calculated for boundary forces equivalent to a temperature difference of $300\,\mathrm{K}$. The off-diagonal component of the strain tensor (the shear strain), being the one coupled to the piezoelectricity, was shown in fig. \ref{fig:fig 4} in the main text. For completeness, the diagonal components of the tensor are shown in fig. \ref{fig:fig s11}. Strain in the $yy$ direction is also expected due to the Poisson effect.

The spatial distribution of $S_{xy}$ agrees qualitatively with analytical half-space solutions for strain induced by surface gates, exhibiting the same characteristic modulation below the gate edges and decay into the substrate. \cite{davies1994theory, murray2006mechanics, hu1991stress}

The diagonal strain components, in particular $S_{xx}$, differ from the corresponding half-space solutions. This difference results from the simplified mechanical boundary conditions used in the finite-element model. The effective forces obtained from the spring model are applied directly as boundary loads and therefore do not reproduce all components of the strain distribution of a fully coupled metal semiconductor half-space. This limitation does not affect the spatial dependence of $S_{xy}$, which is the strain component relevant to the piezoelectric modulation investigated here.

\color{black}


\color{blue}\color{black}The piezoelectric polarization was obtained from the simulated strain tensor using the piezoelectric tensor of GaAs. The different crystal orientations were taken into account by rotating the strain tensor with respect to the axis of the piezoelectric tensor. The resulting  \color{blue}\color{black}electric displacement \color{blue}\color{black}was then used to calculate the electric field and the corresponding electrostatic potential.
\cite{Larkin1997} \color{black}

The piezoelectric back-action onto the strain profile was neglected in this calculation. Other sources of electrostatic potential such as the Schottky barrier at the metal-semiconductor interface or defects were also not included.

\color{blue}\color{black}\subsection{Influence of superlattice periodicity and bar height}\label{influence of superlattice periodicity}

\begin{figure}[htbp]
    \centering
    \includegraphics[width=0.5\textwidth]{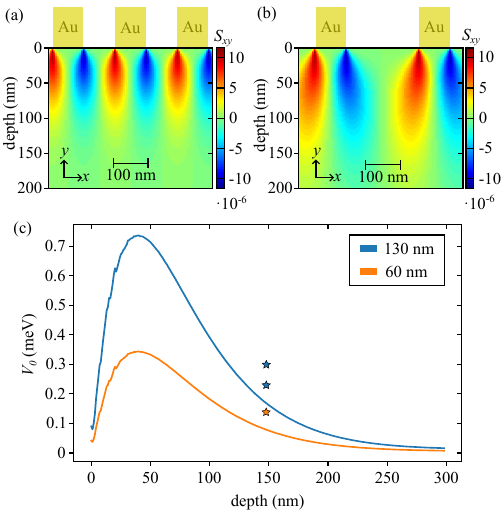}
\caption{\color{blue}\color{black} Calculated heat maps of the shear strain component $S_{xy}$ for LSL periodicity of (a) $180\,\mathrm{nm}$ and (b) $300\,\mathrm{nm}$. A comparison between of the $V_0$ against depth profile is shown in (c) for two different bar heights for the experimentally investigated periodicity $P = 240\,\mathrm{nm}$. The star symbols mark the experimentally observed modulation amplitudes at an effective depth consisting of $125\,\mathrm{nm}$ GaAs and $23\,\mathrm{nm}$ homogeneous Au film. The two blue stars represent the data obtained from the two samples used in fig. \ref{fig:fig 3} and fig. \ref{fig:fig s7}.\\ \color{black}}
\label{fig:fig s12}
\end{figure}

The numerical model was used to investigate the influence of the LSL geometry on the strain-induced potential modulation. In particular the dependence of the LSL periodicity and the LSL bar height was evaluated.
Fig. \ref{fig:fig s12} (a) and (b) present the calculated shear strain profiles for different LSL periodicity. Increasing the distance between neighboring bars increases the depth into the semiconductor before there is an overlap of the strain fields,
allowing the shear strain to propagate deeper into the substrate before balancing out. Consequently, both the magnitude and the depth dependence of $V_0$ are strongly affected by the superlattice periodicity. The corresponding depth dependence of $V_0$ is presented in the main text fig \ref{fig:fig 4} (b).
The dependence of $V_0$ on the 2DEG depth for different 
bar heights is shown in fig \ref{fig:fig s12} (c). Increasing the bar height enhances 
proportionally \color{black} the magnitude of the potential modulation, while the overall shape of the depth dependence is unchanged.

\color{blue}

\FloatBarrier

\end{document}